\documentclass[aps,pra,10pt,showpacs,twocolumn,superscriptaddress,longbibliography]{revtex4-1}
\usepackage{physics}
\usepackage{graphicx}
\usepackage[usenames]{color}
\usepackage{amssymb,amsmath}
\usepackage{siunitx}
\usepackage{xcolor}
\usepackage{setspace} 
\usepackage{float} 
\usepackage{tabularx}
\usepackage[linkcolor=blue,citecolor=blue,colorlinks=true,urlcolor=blue]{hyperref}
\newcommand{\eq}[1]{Eq.~(\ref{#1})}
\newcommand{\fig}[1]{Fig.~\ref{#1}}
\newcommand{\be}[1]{\begin{equation}\label{#1}}
\newcommand{\ee}{\end{equation}}
\usepackage{amsmath,esint}
\usepackage{comment}
\usepackage{lipsum}
\usepackage[toc]{appendix}
\usepackage{hyperref}
\hypersetup{
    colorlinks=false, 
    linktoc=all,     
    linkcolor=blue,  
}

\newcommand\xzplane{$x$\nobreakdash--$z$~plane}

\begin{document}

\title{Mapping the direction of electron ionization to  phase delay between VUV and IR laser pulses}

\author{M. Mountney}
\affiliation{Department of Physics and Astronomy, University College London, Gower Street, London WC1E 6BT, United Kingdom}
\author{G. P. Katsoulis}
\affiliation{Department of Physics and Astronomy, University College London, Gower Street, London WC1E 6BT, United Kingdom}
\author{S. H. M$\o$ller}
\affiliation{Joint Attosecond Science Lab of the National Research Council and the University of Ottawa, Ottawa, Ontario K1A 0R6, Canada}
\author{K. Jana}
\affiliation{Joint Attosecond Science Lab of the National Research Council and the University of Ottawa, Ottawa, Ontario K1A 0R6, Canada}
\author{P. Corkum}
\affiliation{Joint Attosecond Science Lab of the National Research Council and the University of Ottawa, Ottawa, Ontario K1A 0R6, Canada}
\author{A. Emmanouilidou}
\affiliation{Department of Physics and Astronomy, University College London, Gower Street, London WC1E 6BT, United Kingdom}

\begin{abstract}
We theoretically demonstrate a one-to-one mapping between the direction of electron ionization  and the phase delay between a linearly polarized VUV and a circular IR laser pulse. To achieve this, we use an ultrashort VUV pulse that defines the moment in time and space when an above threshold electron is released in the IR pulse. The electron can then be accelerated to high velocities escaping in a direction completely determined by the phase delay between the two pulses. The dipole matrix element to transition from an initial bound state of the N$_2$ molecule, considered in this work, to the continuum is obtained using quantum mechanical techniques that involve computing accurate continuum molecular states. Following release of the electron in the IR pulse, we  evolve classical trajectories, neglecting the Coulomb potential and accounting for quantum interference, to compute the distribution of the direction and magnitude of the final electron momentum. The concept we theoretically develop can be implemented to produce nanoscale ring currents that generate  large magnetic fields.

\end{abstract}

\date{\today}

\maketitle

\section{Introduction}

Generating large and controllable THz magnetic fields is of fundamental  importance. For instance, such fields control the direction of magnetization in magnetic devices and,  when  a maximum speed of magnetic switching is reached, they control the onset of magnetic disorder \cite{ref:Tudosa}. Thus, THz magnetic fields  have a wide range of applications in optoelectronic devices used for information processing. Such fields can currently be produced by the Stanford Linear Accelerator (SLAC) electron beam. However, while important, these fields  are a relatively low-tech  application of a state-of-the-art accelerator. Devising optical techniques for generating THz magnetic fields with sufficient spatial and temporal precision to be useful for small devices is a frontier of ultrafast science.

Recently, utilizing
visible and infrared (IR) laser technology allowed for  high-velocity electrons to be arranged in almost any geometric form in semiconductors \cite{ref:Dupont,ref:Sederberg_Kong, ref:Jana,ref:Sederberg}. Such forms include a line current, as in SLAC, and a ring current, relevant for generating solenoidal magnetic fields. These high-velocity electrons, which  generate high magnetic fields, were produced using coherent control.

Coherent control is a powerful tool with applications in a wide range of areas such as quantum optics and metrology \cite{ref:Garcia,ref:Scharfenberger,ref:Wang}, attosecond metrology \cite{ref:Corkum_Krausz,ref:Boutu}, optoelectronics \cite{ref:Hache} and laser cooling \cite{ref:Viteau,ref:Lien}.
 In recent studies \cite{ref:Dupont,ref:Sederberg_Kong, ref:Jana,ref:Sederberg}, steering the direction of electron current  was achieved by controlling  the quantum interference of excitation or ionization pathways resulting from a mid-IR $\omega$ pulse and its second harmonic 2$\omega$ \cite{ref:Dupont}. The phase delay of  the two pulses  was shown to control interference between the two-photon ($\omega$) and single-photon (2$\omega$) pathways and to finally determine the direction of electron motion \cite{ref:Dupont}. 

However, this optical technique,  limits the dimensions over which the electron current is generated to roughly one wavelength of the infrared light that is used to accelerate the electrons \cite{ref:Jana_Okocha}, i.e. to a few $\si{\micro\metre}$. Using coherent control of one and two photon processes to reduce to the nm scale the dimensions over which current is produced  requires optically generating $\omega$ and 2$\omega$  vacuum ultraviolet (VUV) light beams. This is currently impractical.

Here, we theoretically demonstrate that control of electron currents generated at roughly 100 nm is possible. As in the attosecond streak camera \cite{ref:Constant}, we achieve control by varying the phase delay between a linearly polarized VUV pulse and a circularly polarized IR pulse.  

We show control of electron motion in the context of the N$_{2}$ molecule. Our approach involves, first, releasing the electron at low velocity above the ionization threshold of an atom or molecule using a VUV pulse. The VUV pulse with duration significantly smaller than the IR pulse serves to define the space and time origin of the electron. 
 We consider transitions from an inner or outer valence electron 20--40 eV ($<$100 nm) of N$_{2}$.
Then, the newly released electron is accelerated  by a  circularly polarized IR radiation to speeds that are proportional to the field strength and inversely proportional to frequency. We achieve high electron velocities by choosing the IR pulse to have  long wavelength, $\lambda=2.3$ $\si{\micro\metre}$, and intensity of up to 5$\times$10$^{13}$ W/cm$^2$.

 The current study is theoretical and thus we are not constrained by laboratory restrictions. Hence,  we study N$_{2}$ molecules aligned along the linear polarization of the VUV pulse. 
 We demonstrate a one-to-one mapping between the direction of electron escape and the phase delay between the IR and VUV pulses, achieving  excellent control of electron motion.
 
 To show control of electron dynamics, we develop a hybrid quantum-classical approach. The dipole matrix element to transition from an initial bound state of the N$_2$ molecule to the continuum is obtained using very accurate quantum mechanical techniques. The latter involve computing continuum molecular states.  We obtain these states by solving a system of Hartree--Fock (HF) equations in the single center expansion \cite{ref:Agapi}. Also, we derive in detail and implement in our formulation  the dependence of the dipole matrix element on the angles that determine the direction of electron ionization due to the VUV pulse. 
 
 Following release of the electron in the IR  pulse, we neglect the Coulomb potential and in the context of  the strong field approximation evolve classical trajectories, while fully accounting for quantum interference. The development of the classical aspect of this hybrid approach is based on our previous experience in classical techniques describing ionization in strongly driven systems \cite{ref:AgapiMagnetic}.

\section{Method}

\subsection{Dipole matrix element for an electron ionizing due to a VUV pulse}
In what follows, we outline the derivation of the dipole matrix element for an electron to transition from a bound to a continuum state of a molecule due to the VUV pulse. This matrix element is obtained 
as a function of the excess energy and the direction of ionization of the electron. Both of these parameters will be important for determining the control the IR pulse exercises on electron motion.
\subsubsection{Continuum momentum eigenstates}
First, we formulate the continuum momentum eigenstates of a diatomic molecule in the LAB frame. The $z$ axis of the molecule-fixed (MF) frame is taken to be along the axis of the diatomic molecule. The $z$ axis in the LAB frame is taken along the polarization direction of the linear VUV laser pulse. The origin of the MF and LAB frames is at the center of mass of the molecule. The momentum eigenstate of an electron escaping in the continuum with momentum $\vec{k}$ is expressed in the MF frame (unprimed vectors) \cite{ref:Zare}, \cite{ref:Jacobs} as follows
\begin{equation}
\label{eqn:eigenstateMF}
    \ket{\Vec{k}} = \sum_{l_1=0}^{L}\sum_{m_1=-l_1}^{l_1} i^{l_1}e^{-i\sigma_{l_1}}Y_{l_1,m_1}^*(\hat{k})\psi_{l_1,m_1}(\Vec{r};k),
\end{equation}
where $*$ denotes complex conjugation. 

The basis functions $\psi_{l_1m_1}(\Vec{r};k)$ are the continuum energy eigenstates of a diatomic molecule normalized in energy $\epsilon$, with $\epsilon=\frac{k^2}{2}$. The $\vec{k}$ eigenstate in \eq{eqn:eigenstateMF} is normalized in energy as well. The Coulomb phase shift $\sigma_{l_1}(k)$ is given by $\arg\Gamma(l_1+1-\frac{iZ}{k})$, with $Z$ the net charge of the molecular ion following the escape of one electron. We use atomic units, unless otherwise stated. To obtain in the LAB frame (primed vectors) the continuum momentum eigenstate, we express in the LAB frame the spherical harmonic that is a function of $\hat{k}$ in \eq{eqn:eigenstateMF}. Using the convention that was introduced by Rose in Ref. \cite{ref:Rose} and was also adapted in Refs. \cite{ref:DillDehmer} and \cite{ref:Zare}, we obtain
\begin{equation}
\label{eqn:labtomol}
    Y_{l_1,m_1}(\hat{k}) = \sum_{m_2 = -l_1}^{l_1}Y_{l_1,m_2}(\hat{k}')\mathcal{D}^{l_1}_{m_2,m_1}(\hat{R}).
\end{equation}
The matrix $\mathcal{D}^{l_1}_{m_2,m_1}(\hat{R})$ is the Wigner rotation matrix (see Refs. \cite{ref:Rose} and \cite{ref:Comp}), where $l_1$ is the angular momentum quantum number and $m_1, m_2$ are the magnetic quantum numbers. The three Euler angles $\hat{R} = (\alpha, \beta, \gamma)$ specify the orientation of the diatomic molecule with respect to the LAB frame. 
Next, inverting \eq{eqn:labtomol}, we obtain the spherical harmonics in the LAB frame as follows
\begin{equation}
\label{eqn:moltolab}
    Y_{l_1,m_1}(\hat{k}') = \sum_{m_2}Y_{l_1,m_2}(\hat{k})\mathcal{D}^{l_1*}_{m_1,m_2}(\hat{R}).
\end{equation}
Substituting \eq{eqn:labtomol} in \eq{eqn:eigenstateMF}, the continuum momentum eigenstate in the LAB frame is given by

\begin{equation}
\label{eqn:kprimed}
    \ket{\vec{k}'} = \sum_{l_1,m_1,m_2}i^{l_1}e^{-i\sigma_{l_1}}Y^*_{l_1,m_2}(\hat{k}')\mathcal{D}^{l_1*}_{m_2,m_1}(\hat{R})\psi_{l_1,m_1}(\Vec{r};k).
\end{equation}


\subsubsection{Energy normalized continuum and bound wavefunctions}

To calculate the continuum wavefunctions, we use a single center expansion (SCE)  \cite{ref:Demekhin,ref:Agapi} with respect to the center of mass of the molecule. In SCE, the bound wavefunctions are given by 
\begin{equation}
\label{eqn:BoundSCE}
    \psi_i(\vec{r}) = \sum_{l_i,m_i}\frac{P_{l_im_i}(r)Y_{l_im_i}(\hat{r})}{r},
\end{equation}
where $\vec{r} = (r,\theta,\phi)$. Note that $\psi_i(\vec{r})$ is a HF orbital  in our calculations. The continuum wavefunctions $\psi_{l_1,m_1}(\Vec{r};k)$ are expressed as follows
\begin{equation}
\label{eqn:ContSCE}
    \psi_{l_1,m_1}(\Vec{r};k) = \sum_{l'm'}\frac{\mathcal{P}^{l_1m_1}_{l'm'}(r;k)}{r}Y_{l'm'}(\hat{r}).
\end{equation}

Expressing the wavefunction as a product of radial functions and spherical harmonics significantly simplifies the computation of integrals. Indeed, when solving a system of HF equations, the integrals over angles are computed analytically. As a result, we solve a system of equations expressed in terms of the radial part of both bound and continuum wavefunctions, see Ref. \cite{ref:Agapi}. The energy normalized radial functions $\mathcal{P}^{l_1m_1}_{l'm'}(r;k)$ are found to be \cite{ref:Demekhin,ref:Aberg}:

\begin{equation}
\label{eqn:norm2}
    \mathcal{P}^{l_1m_1}_{l'm'}(r;k) = \sum_{LM}e^{-i\eta_{LM}}\mathcal{U}_{l_1m_1,LM}		\mathcal{P}^{l'm'}_{LM}(r;k),
\end{equation}
where
\begin{equation}
\label{eqn:norm1}
   \mathcal{P}^{l'm'}_{LM}(r;k) = \cos\eta_{LM}\sum_{L'M'}\mathcal{U}_{L'M',LM}\mathcal{P}^{l'm'}_{L'M'}(r;k).
\end{equation} 
The phaseshifts $\eta_{LM}$ and matrix $\mathcal{U}$ are obtained from the following eigenvalue equation
\begin{equation}
    \sum_{L'M'}R_{lm,L'M'}\mathcal{U}_{L'M',LM} = -\tan\eta_{LM}\mathcal{U}_{lm,LM},
\end{equation}
where $\mathcal{U}$ consists of the eigenvectors of the interaction matrix $R$, defined below. The $R$ matrix is referred to as scattering matrix $K$ in Ref. \cite{ref:Zare}. The unnormalized 
continuum radial functions $\mathcal{P}^{l'm'}_{L'M'}(r;k)$ are obtained after solving the system of HF equations.

The energy normalized  functions  $\mathcal{P}^{l_1m_1}_{l'm'}(r;k)$ in \eq{eqn:norm2} result from 
$\mathcal{P}^{l'm'}_{L'M'}(r;k)$ satisfying the normalization condition

\begin{equation}
\label{eqn:boundary2}
\int^{\infty}_0dr\sum_{L'M'}{\mathcal{P}}^{l'm'}_{L'M'}(r;k_1)\mathcal{P}^{l'm'}_{L'M'}(r;k_2) = \delta(\epsilon_1 - \epsilon_2),
\end{equation}
where $\epsilon_1 = \frac{k_1^2}{2}$ and $\epsilon_2 = \frac{k_2^2}{2}$. It follows that the radial functions satisfy the asymptotic conditions \cite{ref:Inhester}
\begin{equation}
\label{eqn:asymptotic}
\begin{split}
\mathcal{P}^{l'm'}_{L'M'}(r\rightarrow \infty;k) &\rightarrow \mathcal{F}_{L'}(r;k)\delta_{l',L'}\delta_{m',M'} 
\\
&+ \mathcal{G}_{L'}(r;k)R_{l'm',L'M'},
\end{split}
\end{equation}
where $\mathcal{F}_{L'}(r;k)$ and $\mathcal{G}_{L'}(r;k)$ are energy normalized regular and irregular Coulomb functions, respectively, see Refs. \cite{ref:Abramowitz,ref:Seaton}. The regular Coulomb function corresponds to the solution of a point charge. The irregular Coulomb function corresponds to the distortion of the solution from that of a single point charge. The latter is present in a molecule since there is no spherical symmetry.

\subsubsection{Dipole matrix element from a bound to a continuum molecular state}\label{section:Dipole}
The photoionization cross-section to transition from an initial  bound state to a final continuum molecular state is proportional to the absolute value squared of the dipole matrix element. The latter describes the transition of an electron from an initial state $\mathcal{\psi}_i(\Vec{r})$ to a final state $\ket{\vec{k}'}$ due to a single photon absorption, here by a VUV pulse, 
\begin{equation}
\label{eqn:dipole1}
    D_M = D_M(\vec{k}') = \mel{\vec{k}'}{\vec{r'}\cdot\hat{n}'}{\psi_i},
\end{equation}
where
\begin{equation}
\label{eqn:dipole2}
   \vec{r'}\cdot\hat{n}' = \sqrt{\frac{4\pi}{3}}rY_{1,M}(\hat{r}'),
\end{equation}
with $\hat{n}'$ being the unit vector of the polarization of the electric field of the VUV pulse. The value of $M$ is 0 for linear polarization and $\pm 1$ for right/left circularly polarized light in the LAB frame. Substituting \eq{eqn:moltolab} into \eq{eqn:dipole2} for $\hat{r}'$ instead of $\hat{k}'$, we obtain
\begin{equation}
\label{eqn:dipole3}
   \vec{r'}\cdot\hat{n}' = \sqrt{\frac{4\pi}{3}}r\sum_{\abs{m}\leq1}Y_{1,m}(\hat{r})\mathcal{D}^{1*}_{M,m}(\hat{R}).
\end{equation}
The polarization of the photon in the MF frame is denoted by $m$. Next, substituting \eq{eqn:dipole3} and \eq{eqn:kprimed} into \eq{eqn:dipole1}, we obtain
\begin{equation}
\label{eqn:DM}
\begin{split}
D_M = \sum_{l_1,m_1,m_2,m}&e^{i\sigma_{l_1}}(-i)^{l_1}\mathcal{D}^{l_1}_{m_2,m_1}(\hat{R})\mathcal{D}^{1*}_{M,m}(\hat{R})
\\
&\times Y_{l_1,m_2}(\hat{k}')D_{l_1,m_1,m},
\end{split}
\end{equation}
where
\begin{equation}
\label{eqn:Dl1m1mi}
    D_{l_1,m_1,m} = \int d\Vec{r} \psi^*_{l_1,m_1}(\Vec{r};k)\sqrt{\frac{4\pi}{3}}rY_{1,m}(\hat{r})\psi_i(\Vec{r}).
\end{equation}

Substituting Eqns. (\ref{eqn:BoundSCE}) and (\ref{eqn:ContSCE}) in \eq{eqn:Dl1m1mi} and integrating over the angular part in \eq{eqn:Dl1m1mi}, we obtain an expression in terms of Wigner-3$j$ symbols and of the radial bound and continuum wavefunctions
\begin{equation}
\begin{split}
\label{eqn:Dl1m1miWigner}
    D_{l_1,m_1,m} &= \sqrt{\frac{4\pi}{3}}\sum_{l',m',l_i,m_i}(-1)^{m'}\sqrt{(2l_i+1)(2l'+1)}
    \\
    &\times\begin{pmatrix}
    l' & l_i & 1\\
    0 & 0 & 0 \\
    \end{pmatrix}
    \begin{pmatrix}
    l' & l_i & 1\\
    -m' & m_i & m \\
    \end{pmatrix}
    \\
    &\times
    \int^{\infty}_0 dr \mathcal{P}^{l_1m_1}_{l'm'}(r)rP_{l_im_i}(r).
\end{split}
\end{equation}

 For diatomic molecules, the magnetic quantum number is a good number, with no summation over the initial state quantum number $m_i$ in \eq{eqn:Dl1m1miWigner} and over $m_1$ in \eq{eqn:DM}. Also, from the properties of the Wigner-3$j$ symbol, it follows that in \eq{eqn:Dl1m1miWigner} $m' = m + m_i$. Hence, there is no summation over $m'$ in \eq{eqn:Dl1m1miWigner}. Given the above, \eq{eqn:DM} takes the form:
\begin{equation}
\label{eqn:DM_Diatomic}
\begin{split}
D_M = \sum_{l_1,m_2,m} & e^{i\sigma_{l_1}}(-i)^{l_1}\mathcal{D}^{l_1}_{m_2,m + m_i}(\hat{R})\mathcal{D}^{1*}_{M,m}(\hat{R})
\\
&\times Y_{l_1,m_2}(\hat{k}')D_{l_1,m + m_i,m}.
\end{split}
\end{equation}

Here, we consider  the symmetry axis of the diatomic molecule being parallel to the VUV pulse, with the latter being polarized along the  $z$ axis in the LAB frame. This corresponds to Euler angles $\alpha=\beta=\gamma=0$, resulting in $\mathcal{D}^{l_1}_{m_2,m + m_i}(\hat{0}) = \delta_{m_2, m + m_i}$ and $\mathcal{D}^{1*}_{M,m}(\hat{0}) = \delta_{M, m}$, i.e. $M=m$ and $m_2 = M + m_i$. Given the above, the dipole matrix element for a diatomic molecule aligned along the polarization direction of the VUV pulse takes the form
\begin{equation}
\label{eqn:DM_Diatomic_Parallel}
D_M = \sum_{l_1}e^{i\sigma_{l_1}}(-i)^{l_1}Y_{l_1,M+m_i}(\hat{k}')D_{l_1,M+m_i,M}.
\end{equation}

\subsection{Transition amplitude of an electron from a bound to a continuum state  due to VUV+IR pulses}

 We have obtained the dipole matrix element $D_M(\vec{k}')$ for an electron to transition from a bound molecular state with ionization energy equal to $I_p$ to a final continuum molecular state $\vec{k}'$ with $\frac{\Vec{k}'^2}{2}=\hbar\omega - I_p$, where $\hbar\omega$ is the photon energy of the VUV pulse. Next, in terms of this dipole matrix element, we obtain the amplitude to transition from an initial bound state of the molecule to a final continuum state with electron momentum $\vec{p}_{f}$ in the presence of the VUV and IR pulses.

After an electron is released into the continuum with momentum $\vec{k}'$ at time $t_{\text{ion}}$ by the VUV pulse, we neglect the Coulomb potential. The duration of the VUV pulse is taken to be much smaller than the duration of the IR pulse. The electron is then accelerated by a circular IR laser pulse polarized on the \xzplane. Hence, the conserved canonical momentum due to the motion of the electron in the IR laser field is given by
\begin{equation}
\label{eqn:pfinal}
    \vec{k}'(t_{\text{ion}}) - \Vec{A}_{\text{IR}}(t_{\text{ion}}) = \vec{k}'(t) - \Vec{A}_{\text{IR}}(t) = \Vec{p}_f,
\end{equation}
where $\vec{p}_f$ is the final electron momentum at the end of the IR laser field.

 The envelope of the electric field of the VUV pulse that ionizes the electron is given by
\begin{equation}
    \Vec{E}_{\text{V}}(t) = E_0^{\text{V}}\exp\left[ -2\log(2)\left( \frac{t}{\tau_{\text{V}}} \right)^2\right]\hat{z},
\end{equation}
where $E_0^{\text{V}}$ is the amplitude and $\tau_{\text{V}}$ is the full width at half maximum (FWHM) in intensity of the VUV pulse.
The vector potential of the IR pulse is given by

\begin{align}
\label{eqn:IRVecPot}
\begin{split}
&\Vec{A}_{\text{IR}}(t) = -\frac{E_0^{\text{IR}}}{\omega_{\text{IR}}}
    \exp\left[-2\log(2)\left(\frac{t+\Delta t}{\tau_{\text{IR}}}\right)^2\right] \\
&\times\left\{ \sin\left[ \omega_{\text{IR}}( t + \Delta t) \right] \hat{x} + \cos\left[ \omega_{\text{IR}}( t + \Delta t) \right] \hat{z}  \right\}
\end{split}
\end{align}
where $E_0^{\text{IR}}$ is the amplitude and $\omega_{\text{IR}}$ is the frequency of the electric field of the IR pulse, with $\tau_{\text{IR}}$ being the FWHM in intensity. In addition, $\Delta t$ is the time delay between the VUV and IR pulses. In what follows, we refer to $\omega_{\text{IR}} \Delta t$ as phase delay $\phi$.

According to the strong field approximation (SFA), \cite{ref:Smirnova,ref:CDLin}, the amplitude for an electron to transition from a bound state $\psi_i$ to a final state with momentum $\vec{p}_f$ in the presence of the XUV+IR laser fields is given by 
\begin{equation}
\label{eqn:ampfinal}
\begin{split}
    a(\vec{p}_f) = \int_{t_{i}}^{t_f}&dt_{\text{ion}}E_{\text{V}}(t_{\text{ion}})D_M(\Vec{p}_f+\Vec{A}_{\text{IR}}(t_{\text{ion}}))
    \\
    &\times e^{-iS(t_{\text{ion}},t_f,\vec{p}_f)},
\end{split}
\end{equation}
The times $t_{i}$ and $t_f$ denote the start and end, respectively, of the IR laser field. The classical action $S$ accumulated during the time interval from  $t_{\text{ion}}$ until $t_{f}$ is given by
\begin{equation}
\label{eqn:Action}
\begin{split}
    S(t_{ion},t_f,\vec{p}_f) &= - I_pt_{\text{ion}} +  \int^{t_f}_{t_{\text{ion}}}dt'\frac{ \left[ \Vec{p}_f+\Vec{A}_{\text{IR}}(t') \right]^2}{2}
    \\
    &= \dfrac{ p_f^2}{2} (t_f-t_{\text{ion}})-I_pt_{\text{ion}}
    \\
    &+\int_{t_{\text{ion}}}^{t_f} dt \dfrac{\Vec{A}_{\text{IR}}(t) \cdot \left[ \Vec{A}_{\text{IR}}(t) + 2 \vec{p}_f \right]  }{2} .
\end{split}
\end{equation}

Next, we describe how to compute $a(\vec{p}_f)$ in \eq{eqn:ampfinal} for each photon energy of the VUV pulse. First, we create a two-dimensional grid over the polar angle $\theta_{\text{V}}$ and azimuthal angle $\phi_{\text{V}}$ that define the direction of ejection of the escaping electron with momentum $\vec{k}'$ due to the VUV pulse. For each point of the two-dimensional grid, we compute fully quantum mechanically $D_M(\vec{k}')$, as described in section \ref{section:Dipole}. Then, we describe classically the propagation in the IR pulse of the electron ejected with momentum $\vec{k}'(t_{\text{ion}})$. Specifically, we choose the ionization time $t_{\text{ion}}$ using importance sampling \cite{ref:ImpSamp} in the time interval $[-2.5\tau_{\text{V}},2.5\tau_{\text{V}}]$. For the probability distribution, we use the amplitude of the VUV pulse $E^{\text{V}}_0$. For each classical trajectory, we propagate the electron in the IR laser field from time $t_{\text{ion}}$ to time $t_f$. We generate, for each $\theta_{\text{V}}$ grid point, $2\times 10^7$ classical trajectories.

 In addition, we account for the interference of 
trajectories corresponding to electrons ejected with different momenta $\vec{k}'$ at different ionization times $t_{\text{ion}}$ that finally escape with the same momentum $\vec{p}_f$. To do so, we create a three-dimensional grid over the cylindrical coordinates ${p_f}_r, 
\theta,{p_f}_y$ of the final momentum that the electron acquires due to both the VUV and IR pulses. We note that $p{_f{_r}}$ is the magnitude of the projection of $p_{f}$ on the plane of the IR pulse, i.e. $p{_f{_r}}=\sqrt{p{_f{_x}}^2+p{_f{_z}}^2}$. Also, $\theta$ is the angle of $p{_f{_r}}$ with the $z$ axis, which is also the axis of 
polarization of the VUV pulse. We obtain the amplitude $\mathcal{A}(\vec{p}_f)$ for each grid point corresponding to an electron momentum $\vec{p}_f$ by adding coherently the amplitudes $a_i$ for all trajectories $i$ with the 
same $\vec{p}_f$ as follows
\begin{equation}
\abs{\mathcal{A}(\vec{p}_f)}^2 = \abs{\sum_{i}a_i(\vec{p}_f)}^2.
\end{equation}
Finally, we obtain the probability for an electron to be ejected on the plane \xzplane \ of the IR pulse with momentum $(p{_f{_r}}, \theta)$ by integrating $\abs{\mathcal{A}(\vec{p}_f)}^2$ over the ${p_f}_y$ component 
\begin{equation}
\label{eq:projection}
    \abs{\mathcal{A}(p{_f{_r}}, \theta)}^2 = \int dp{_f{_y}} \abs{\mathcal{A}(\vec{p}_f)}^2.
\end{equation}

\section{Results}




Next, we demonstrate that by changing the phase delay $\phi$ between the VUV and IR laser pulses, we control the direction of escape of an electron that is released in the continuum due to the VUV pulse and accelerates due to the IR pulse. We do so in the context of   N$_2$, when the VUV pulse is aligned with the $z$ axis in the LAB frame. \subsection{Computation of the bound and continuum orbitals}

 We  briefly discuss the computation of the bound and continuum orbitals of N$_2$.
The electronic configuration of N$_2$ is ($1\sigma_g^2$, $1\sigma_u^2$, $2\sigma_g^2$, $2\sigma_u^2$, $3\sigma_g^2$, $1\pi_{ux}^2$, $1\pi_{uy}^2$). We consider single photon absorption by the VUV pulse and subsequent ionization of an electron initially occupying the $2\sigma_g$, $3\sigma_g$ or $1\pi_u$ bound orbital. The $1\pi_{ux}$ and $1\pi_{uy}$ orbitals are energy degenerate and have opposite $m$ quantum number. For  the computations performed in this work, the initial state $\psi_i$ is given by one of these three orbitals. We compute the bound states using the HF method with the quantum chemistry package MOLPRO \cite{ref:Molpro}. We implement HF by employing the correlation-consistent polarized triple-zeta basis set (cc-pVTZ) \cite{ref:Dunning}. We find the equilibrium distance to be equal to $2.08$ \AA \ and the ionization energies of the $2\sigma_g$, $3\sigma_g$ and $1\pi_u$ orbitals to be equal to 37.7 eV, 16.0 eV and 15.3 eV, respectively,  close to the experimental values reported  in \cite{ref:Cacelli}. 

In the single center expansion of the bound states $\psi_i$, see \eq{eqn:BoundSCE}, it suffices to truncate the expansion over the $l_i$ quantum number up to $l_{\text{max}} = 30$. For the computation of the continuum orbitals $\psi_{l_1,m_1}$, in the single center expansion given in \eqref{eqn:ContSCE} we truncate the $l_1$ quantum number up to $l_{\text{max}} = 19$. We also checked (not shown) that our results for the total cross sections as a function of photon energy for the $2\sigma_g$, $3\sigma_g$ and $1\pi_u$ orbitals are in agreement with experimental results \cite{ref:Samson,ref:Woodruff,ref:Hamnett}.

We take the amplitude and duration of the electric field of the VUV pulse to be equal to $E_0^{\text{V}}=10^{13}$ W/cm$^2$ and $\tau_{\text{V}}=0.5$ fs. We consider two amplitudes of the IR pulse corresponding to intensity of either 5$\times 10^{12}$ W/cm$^2$ or 5$\times 10^{13}$ W/cm$^2$, to roughly identify the strength of the IR pulse required to achieve control. Given the ionization energy of a valence electron of N$_2$, we find that the intensity upper limit of 5$\times 10^{13}$ W/cm$^2$  results by requiring that the rate of ionization of a valence electron via tunnelling due to the IR pulse is very small. 

In our studies, the duration of the IR pulse is $\tau_{\text{IR}} = 100$ fs and its frequency is equal to $\omega_{\text{IR}} = 0.020$ a.u. (2300 nm). In the two-dimensional grid of the angles of ejection of the electron due to the VUV pulse in the LAB frame, the polar angle $\theta_{\text{V}}$ ranges from 0$^{\circ}$ to $180^{\circ}$ in steps of 1$^{\circ}$, while the azimuthal angle ranges $\phi_{\text{V}}$ from 0$^{\circ}$ to $360^{\circ}$ in steps of $10^{\circ}$. For the three-dimensional grid of the final electron momentum in cylindrical coordinates, p${_f{_y}}$ and p${_f{_r}}$ vary from $-5$ a.u. to 5 a.u. and 0 a.u. to 5 a.u., respectively, in steps of 0.01 a.u., while the angle $\theta$ varies from 0$^{\circ}$ to 360$^{\circ}$ in steps of 1$^{\circ}$.  

\subsection{Control of direction of electron ionization with VUV+IR pulses}
We demonstrate that the phase delay between the linearly polarized VUV pulse and the circularly polarized IR pulse determines the direction of the ionizing electron. We show that best control is achieved when, at the time $t_{\text{ion}}$ that the electron is released by the VUV pulse, it has very small energy. That is, the photon energy of the VUV pulse has to be just above the ionization threshold for the IR pulse to steer the electron most effectively.

To illustrate this, we plot the probability $g(\theta)$ for an electron to escape to the continuum on the \xzplane \ of the IR pulse with an angle $\theta$. The latter angle is measured with respect to the $z$ axis in the LAB frame.  Integrating $\abs{\mathcal{A}(p{_f{_r}}, \theta)}^2$ in \eq{eq:projection} over  $p{_f{_r}}$, we find g($\theta$) as follows

\begin{equation}
\label{eqn:ftheta}
    g(\theta) = \int dp_{f_r} p_{f_r}\abs{\mathcal{A}(\vec{p}_f,\theta)}^2.
\end{equation}

  In our results, we fully account for the energy range of the VUV pulse. That is, the Fourier transform of the VUV pulse with a full-width half maximum equal to 0.5 fs extends over energies roughly $\pm$ 4 eV from the central 
photon energy. Hence, when considering a VUV pulse with central photon energy equal to, for instance, 23 eV we also consider photon energies in the interval [19, 27] eV in steps of 1 eV. That is, for each of these energies we obtain the amplitudes $a_{i}(\vec{p}_f)$ which we then weight by the value of the Fourier transform at the respective energy. We then add coherently the amplitudes thus obtained  to compute $\abs{\mathcal{A}(p{_f{_r}}, \theta)}^2$ in \eq{eq:projection}.

\begin{figure*}
    		\includegraphics[width=1\textwidth]{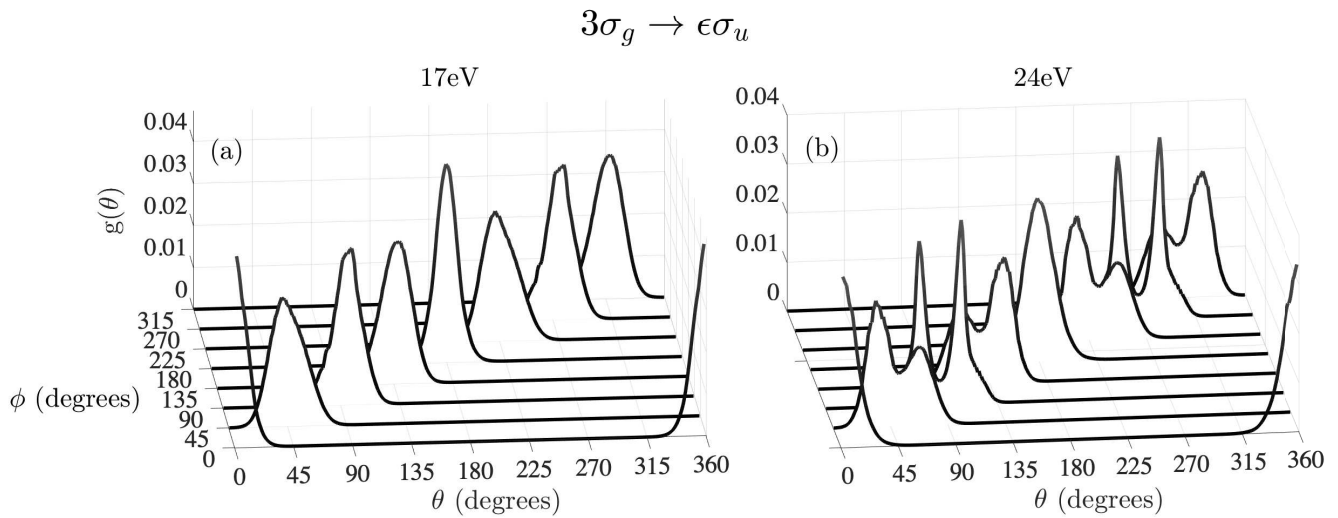}
    		\caption{Probability $g(\theta)$ for an electron ionizing from the 3$\sigma_{g}$ orbital due to a linearly polarized VUV pulse and accelerated by a circularly polarized IR pulse to ionize with an angle $\theta$ with respect to the $z$ axis on the $x$--$z$ plane. The IR pulse is polarized on the $x$--$z$ plane. The VUV pulse is linearly polarized along the molecular axis ($z$ axis). We waterfall plot the probability $g(\theta)$ for different phase delays $\phi$ between the VUV and IR pulses. The VUV photon energy is 17 eV (a) and 24 eV (b). The intensity of the IR pulse is 5$\times 10^{13}$ W/cm$^2$.}
			\label{fig:3s}
\end{figure*}

  In \fig{fig:3s}, we plot the probability distribution $g(\theta$) when an electron ionizes to the continuum from 3$\sigma_{g}$, an outer valence orbital,  with an angle $\theta$ with respect to the $z$ axis on the \xzplane . Since we consider linear polarization of the VUV pulse which is aligned with the molecular axis, the relevant transition is 3$\sigma_{g}\rightarrow\epsilon\sigma_{u}$. The $z$ axis is both the symmetry axis of the diatomic molecule and the polarization axis of the VUV pulse. The circular IR pulse is polarized on the \xzplane . We take the IR pulse to have an intensity of 5$\times 10^{13}$ W/cm$^2$. We consider a 17 eV VUV photon energy, which is 1 eV above the ionization threshold of the $3\sigma_{g}$ orbital (\fig{fig:3s}(a)), and a higher photon energy of 24 eV (\fig{fig:3s}(b)).
  
  In \fig{fig:3s}, we vary the phase delay $\phi$ between the VUV and IR pulses from $0^{\circ}$ to $315^{\circ}$ in steps of 45$^{\circ}$. For each phase delay $\phi$, we find the angle $\theta$ that corresponds to the maximum of $g(\theta$), i.e. the most probable angle of electron escape on the \xzplane , $\theta_{\text{max}}$. For each $\phi$, we expect that $\theta_{\text{max}}=\phi$; we find this to be true when the electron due to the VUV pulse is released at $t_{\text{ion}}$ in the IR pulse with very small excess energy. That is $k'(t_{\text{ion}})=\sqrt{2(\hbar\omega-I_{p})}$ is very small. This is seen in \fig{fig:3s} where $g(\theta$) is narrow and centered around $\theta=\theta_{\text{max}}=\phi$ when the excess energy is 1 eV (17 eV photon energy), while $g(\theta$) is wide and in most cases doubly-peaked for 8 eV excess energy (24 eV photon energy). This finding means that we achieve a one-to-one mapping between the angle of ionization of the electron and the phase delay between the two laser pulses. This holds true for small electron energies at the time the electron is released in the IR pulse. 
  
Hence, we clearly demonstrate control of electron motion for VUV energies around 20 eV. This is important because these energies are within the range of high-harmonics generated from solids where one can harness the surface structure that can be added to solids to focus the VUV light down to the 100 nm scale \cite{ref:Korobenko}.

  In \fig{fig:3s}, we find that when the electron due to the VUV pulse is released at $t_{\text{ion}}$ in the IR pulse with higher excess energy, there is a double peak structure in g($\theta$), mostly for $\phi=90^{\circ}, 270^{\circ}$. For instance for $\phi=90^{\circ}$, the momentum that the electron gains from the IR pulse is equal to $-\vec{A}_{\text{IR}}(t_{\text{ion}})$ and points along the $x$ axis. In addition, the electron is released due to the VUV pulse at time $t_{\text{ion}}$ in the IR pulse  along the $+z$ and $-z$ axis with the same probability, since N$_{2}$ is a homonuclear molecule. As a result, the final angle of electron escape is smaller than 90$^{\circ}$ for electrons released along the $+z$ axis and greater than 90$^{\circ}$ for electrons released  along the $-z$ axis.  
  
\begin{figure*}
	\includegraphics[width=1\textwidth]{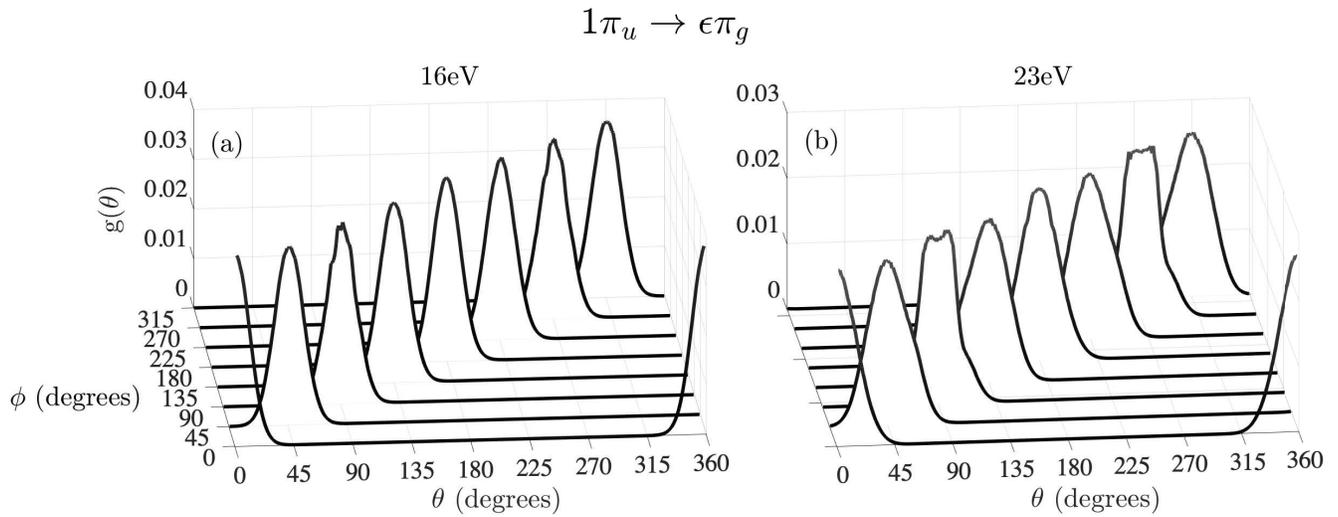}
	\caption{Same as \fig{fig:3s} but for an electron ionizing from the 1$\pi_{u}$ orbital. The VUV photon energy is 18 eV (a) and 23 eV (b). }
	\label{fig:1px}
\end{figure*}

\begin{figure*}
	\includegraphics[width=1\textwidth]{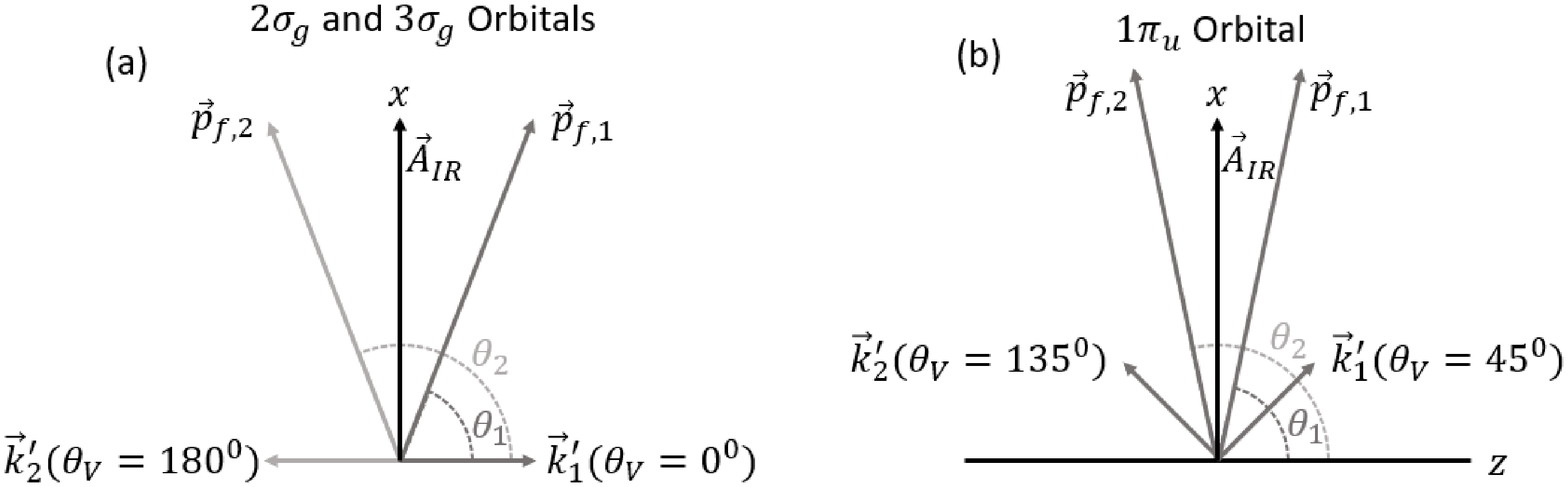}
	\caption{Schematic diagram for the resultant final momentum $\Vec{p}_f$. The vectors $\vec{k}_1'$ and $\vec{k}_2'$ are the momentum vectors due to the VUV pulse corresponding to the two most probable angles of release due to VUV pulse at time $t_{\text{ion}}$. The resultant vectors $\vec{p}_{f,1}$ and $\vec{p}_{f,2}$ are obtained by $\vec{k}_1'-\vec{A}_{\text{IR}}$ and $\vec{k}_2'-\vec{A}_{\text{IR}}$, respectively. The angles $\theta_1$ and $\theta_2$ are the polar angles on the $x$--$z$ plane of the vectors $\vec{p}_{f,1}$ and $\vec{p}_{f,2}$. The diagram on the left corresponds to the $2\sigma_g$ and $3\sigma_g$ orbitals and on the right to the 1$\pi_{u}$ orbital.}
	\label{fig:VecAddition}
\end{figure*}

  Similar results are obtained for ionization to the continuum of an electron from 1$\pi_{u}$, another outer valence orbital, see \fig{fig:1px}. In this case, we achieve excellent control of the angle of electron escape both for 18 eV photon energy (2.7 eV excess energy) but also for a higher photon energy of 23 eV (7.7 eV excess energy). We note that 23 eV photons are feasible for solid state high-harmonic generation. In what follows, we explain the reason for achieving for higher excess energies better control for the 1$\pi_{u}$ orbital compared to the $2\sigma_g$ and $3\sigma_g$ orbitals. We find that the most probable angles of release due only to the VUV pulse are $45^{\circ}$ and $135^{\circ}$ for the 1$\pi_{u}$ orbital, while they are $0^{\circ}$ and $180^{\circ}$ for the $2\sigma_g$ and $3\sigma_g$ orbitals. We refer to the $\vec{k}'$ vectors corresponding to these two angles as $\vec{k}_1'$ and $\vec{k}_2'$. To account for the IR pulse, we add to $\vec{k}_1'$ and $\vec{k}_2'$ the $-\vec{A}_{\text{IR}}$ vector. This addition is illustrated in \fig{fig:VecAddition}(a) for the $2\sigma_g$ and $3\sigma_g$ orbitals and \fig{fig:VecAddition}(b) for the 1$\pi_{u}$ orbital. Comparing \fig{fig:VecAddition}(a) with \fig{fig:VecAddition}(b), it is clear that the angle between the two resultant vectors $\vec{p}_{f,1}$, $\vec{p}_{f,2}$, which is given by $\theta_2 - \theta_1$, is smaller for the 1$\pi_{u}$ orbital than for the $2\sigma_g$ and $3\sigma_g$ orbitals. This is the reason that the double peak structure is significantly less pronounced for the 1$\pi_{u}$ orbital compared to the $2\sigma_g$ and $3\sigma_g$ orbitals for higher excess energies. Indeed, this can be seen by comparing \fig{fig:1px}(b) with \fig{fig:3s}(b) for $\phi=90^{\circ}$ and $\phi=270^{\circ}$.

\begin{figure}
\includegraphics[width=0.5\textwidth]{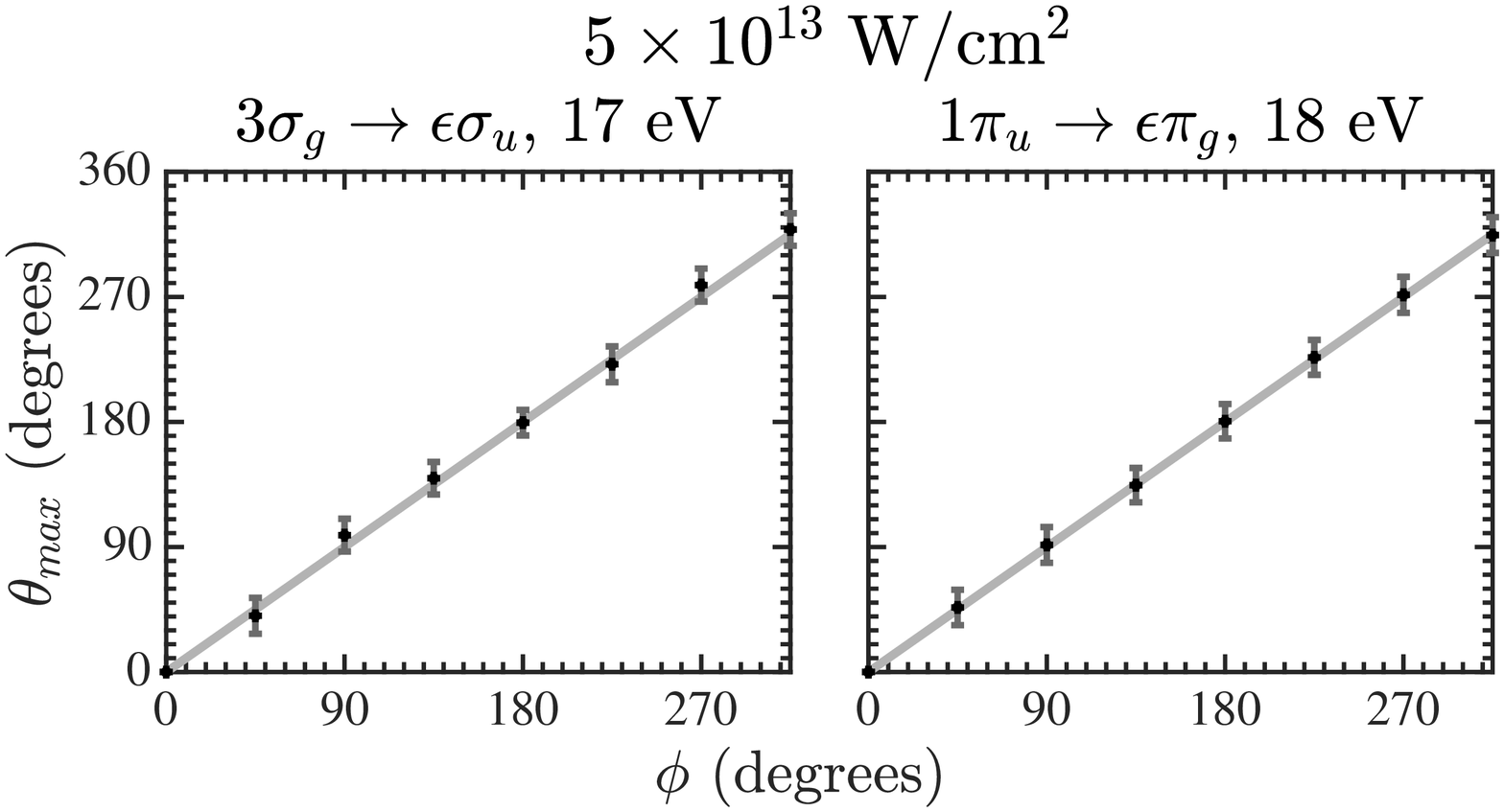}
    \caption{$\theta_{\text{max}}$ as a function of the  delay $\phi$ between the VUV and IR pulses, for an electron ionizing from the 3$\sigma_{g}$ and 1$\pi_{u}$ orbital with VUV photon energy equal to 17 eV and 18 eV, respectively. The intensity of the IR pulse is 5$\times 10^{13}$ W/cm$^2$. The vertical bars denote the standard deviation of the probability distribution g($\theta$).}
    \label{fig:3s_theta_v_phi}
\end{figure}

We further illustrate the one-to-one mapping between the most probable angle of ejection of the electron, $\theta_{\text{max}}$, and the phase delay between the VUV and IR pulses by plotting $\theta_{\text{max}}$ as a function of $\phi$ in \fig{fig:3s_theta_v_phi}. We do so for a transition from the 3$\sigma_{g}$ and $1\pi_{u}$ orbitals when the VUV photon energy is close to the ionization threshold, 17 eV and 18 eV, respectively. The values of $\theta_{\text{max}}$ for all $\phi$s lie on the grey line in \fig{fig:3s_theta_v_phi} that corresponds to $\theta_{\text{max}}=\phi$. For each $\phi$, we also compute the standard deviation of the probability distribution $g(\theta$) and find it to be very small, see \fig{fig:3s_theta_v_phi}. The small spread of the angles $\theta$ around $\phi$, for each $\phi$, implies excellent control of electron motion.
  
  As we have already noted, the Coulomb potential is fully accounted for the interaction of the N$_2$ molecule with the VUV pulse. We neglect the Coulomb potential only during the propagation inside the IR pulse of the electron released at time $t_{\text{ion}}$. We expect that this approximation will not affect our finding of the one-to-one mapping between the phase delay $\phi$ and the most probable angle of electron escape $\theta_{\text{max}}$. Fully accounting for the Coulomb potential at all stages will most probably result in a broader distribution $g(\theta)$ that still has a peak around $\theta_{\text{max}}=\phi$. Moreover, the details of the double peak structure of $g(\theta)$ observed for higher photon energies, see \fig{fig:3s}(b) for $\phi=90^{\circ}$ and $\phi=270^{\circ}$, depend on the system that is interacting with the VUV and IR pulses. For instance, a more asymmetric double peak structure of $g(\theta)$ is expected for a heteronuclear diatomic molecule. The reason is that the probability  for the electron to be released due to the VUV pulse is different when the electron escapes along one centre versus the other one. Finally, we note that in this work we assume that the diatomic molecule is perfectly aligned with the VUV laser pulse. In an experiment, the molecule will be aligned along the VUV pulse with a certain distribution. Accounting for such a distribution is expected to only slightly change the results presented in this work and to cause a small increase of the width of the distribution of g($\theta$) as a function of $\theta$ for different phase delays. This is the reason we do not include   a distribution of alignments in the current work.
  
\begin{figure*}
    \includegraphics[width=1\textwidth]{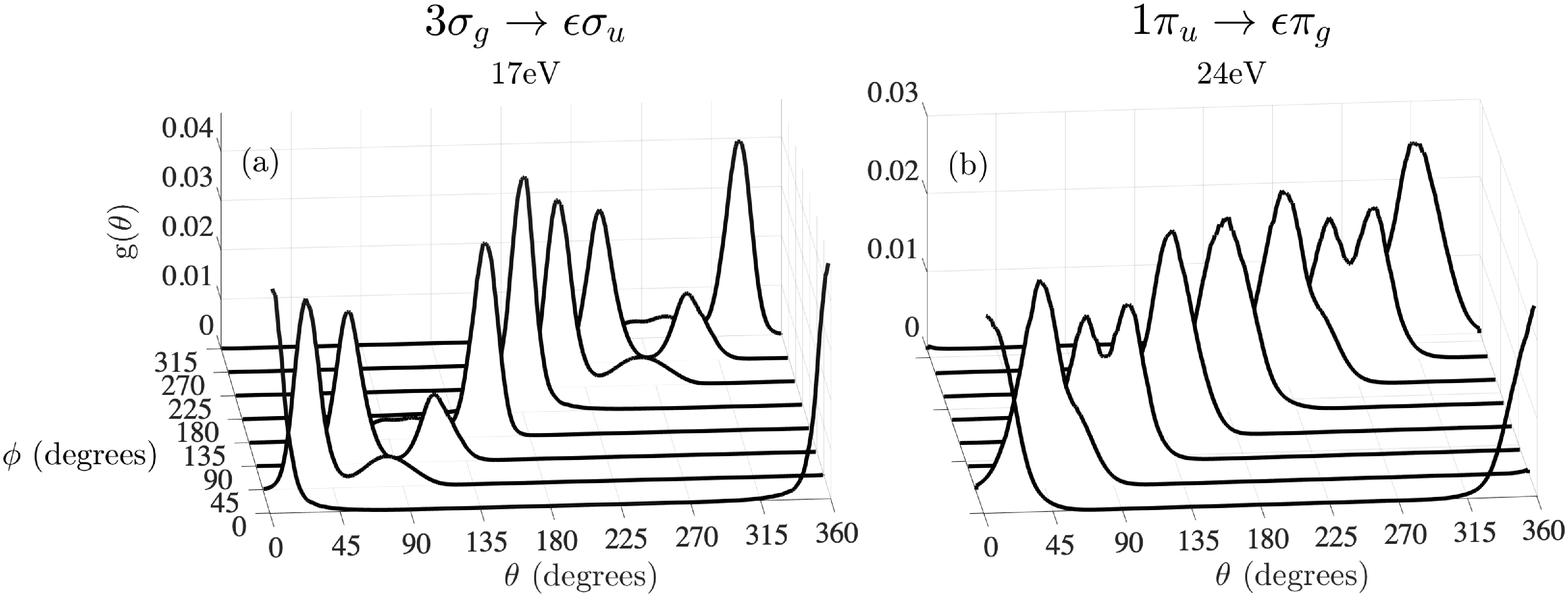}
    \caption{Same as in \fig{fig:3s} and \fig{fig:1px} but for an intensity of the IR pulse equal to 5$\times 10^{12}$ W/cm$^2$.}
    \label{fig:33s}
\end{figure*}

In \fig{fig:33s}, we demonstrate that, as expected, control of the angle of ionization of the electron depends on the strength of the IR pulse. We take the intensity of the IR pulse to be equal to 5$\times 10^{12}$ W/cm$^2$, which is an order of magnitude smaller than the intensity of the IR pulse considered in \fig{fig:3s} and \fig{fig:1px}. We find that for the weaker IR pulse, the distribution g($\theta$) is wide and not centered around $\theta=\phi$ even for VUV photon energies close to the ionization threshold,  18 eV and 17 eV for the transitions from the 1$\pi_{u}$ and  3$\sigma_{g}$ orbital, respectively. The lack of control is clearly seen by the doubly peaked structure of $g(\theta$) for most $\phi$ for both transitions. 

\begin{figure*}
    \includegraphics[width=1.\textwidth]{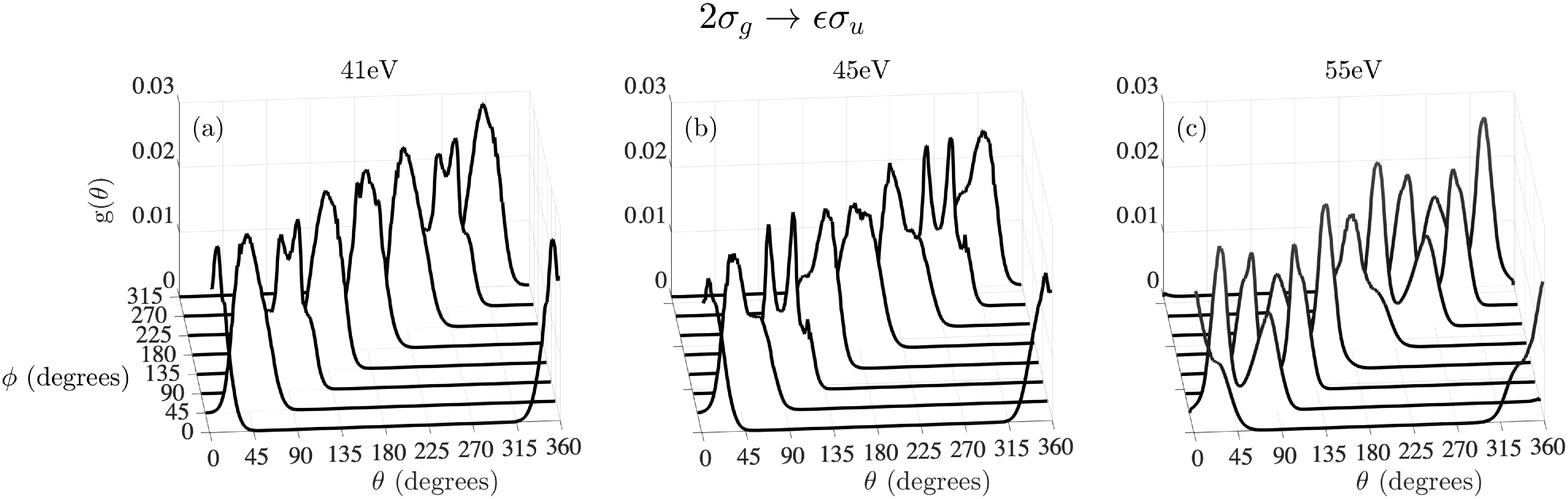}
    \caption{Same as in \fig{fig:3s} but for an electron ionizing from the 2$\sigma_{g}$ orbital. The waterfall plots (a), (b) and (c) correspond to VUV photon energies of 41 eV, 45 eV and 55 eV respectively.} 
   \label{fig:2s}
\end{figure*}

We now show that control of the angle of electron ionization is also achieved for electrons ionizing due to the VUV pulse from the 2$\sigma_{g}$ inner valence  orbital. This is demonstrated in \fig{fig:2s}(a) for photon energy of the VUV pulse equal to 41 eV, which corresponds to the electron having an excess energy of 3.3 eV when it is released into the IR pulse. However, for photon energies 45 eV and 55 eV, as for the 3$\sigma$ orbital at 24 eV VUV photon energy, we find that the distribution $g(\theta$) is wide and doubly-peaked for most values of the phase delay $\phi$, see \fig{fig:2s}(b) and \fig{fig:2s}(c). Hence, control is not achieved for these higher excess energies.

Next, we investigate the velocities reached by the electron that is  released  due to the VUV pulse and is accelerated by the circular IR pulse. In \fig{fig:1s_p}, we plot the distribution of momenta $g(p_{f_{r}}$) the electron escapes with on the \xzplane , which is the plane of polarization of the IR pulse. We do so for an electron ionizing from the 3$\sigma_{g}$ orbital for a 17 eV photon energy of the VUV pulse. The distribution of the  magnitude of the projection of the final electron momentum  on the \xzplane \ is given by 

\begin{equation}
\label{eqn:fp}
    g(p_{f_{r}}) = \int d\theta p_{f_{r}}\abs{\mathcal{A}(\vec{p}_f,\theta)}^2.
\end{equation}
The maximum momentum an electron gains from the IR field is $E_{0}^{\text{IR}}/\omega$, which is equal to roughly 1.9 a.u. for the higher IR intensity considered in this work. The momentum of the electron at the time of release  $t_{\text{ion}}$ is $k'=0.27$ a.u. Hence, when the phase delay $\phi$ between the two pulses is zero, an electron released due to the VUV pulse along the $+z$ axis  at time $t_{\text{ion}}$ 
will finally ionize with a momentum equal to 0.27+1.9=+2.17 a.u. due to both the VUV and IR pulses. An electron released due to the VUV pulse along the $-z$ axis  at time $t_{\text{ion}}$ 
will finally escape with a momentum equal to -0.27+1.9=1.63 a.u. due to both the VUV and IR pulses. This explains the doubly-peaked distribution of electron momenta in \fig{fig:1s_p}, when $\phi=0$ and for most $\phi$s. The height of both peaks is roughly equal, since N$_{2}$ is a homonuclear molecule and the electron has the same probability to be released along the $+z$ and $-z$ axis  at time $t_{\text{ion}}$. Moreover, when $\phi=90^{\circ}$, 
$-\vec{A}_{\text{IR}}(t_{\text{ion}})$ points along the $x$ axis and hence the resultant electron momentum distribution due to the VUV and IR pulses is centered around $\vec{A}_{\text{IR}}(t_{\text{ion}})=E_{0}^{\text{IR}}/\omega_{\text{IR}}=1.9$ a.u. Thus, we find that the ionizing electron is  steered to a specific direction by the phase delay of the VUV and IR pulses 
but it also accelerates to high velocities which are roughly equal to 2$\times 10^{6}$ m/s. 
 If we take an even smaller photon energy of the VUV pulse,  the electron momentum distribution (not shown) will be centered around $E_{0}^{\text{IR}}/\omega_{\text{IR}}$ for most $\phi$s. 

\begin{widetext}
    \begin{minipage}{\linewidth}
        \begin{figure}[H]
    		\includegraphics[width=1\textwidth]{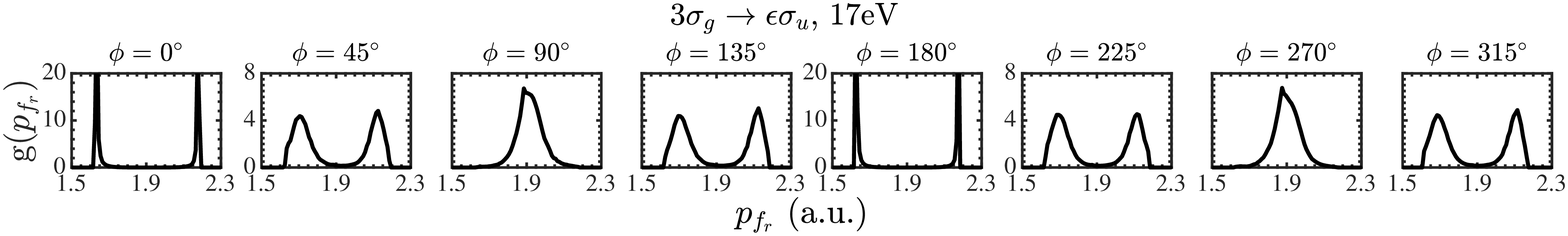}
   	 		\caption{Probability for an electron being released from the 3$\sigma_{g}$ orbital due to a linearly polarized VUV pulse and being accelerated  by a circularly polarized IR pulse to a final momentum whose projection on the $x$--$z$ plane has magnitude $p{_f{_r}}=\sqrt{p{_f{_x}}^2+p{_f{_z}}^2}$. The VUV photon energy is 17eV and the intensity of the IR pulse is 5$\times 10^{13}$ W/cm$^2$.}
    		\label{fig:1s_p}
		\end{figure}    
    \end{minipage}
\end{widetext}

\subsection{Magnetic field due to a ring current}

Finally, we give a rough estimate for the magnetic field resulting from a ring current generated by a linear VUV pulse of intensity 10$^{13}$ W/cm$^2$ and pulse duration of 0.5 fs as well as a circular IR field 
of intensity 5$\times 10^{13}$ W/cm$^2$. For a VUV pulse focused down to roughly 100 nm with 20 eV photon energy the number of photons provided in an area of 100$^2$ nm$^2$ is roughly 2$\times$10$^{5}$. Then, the ionized atoms and thus the electrons released by the VUV pulse in the IR pulse is roughly 2$\times$10$^{5}$. We showed  that an electron released by a VUV pulse in an IR pulse accelerates by the IR field to speeds $\approx$2$\times 10^{6}$ m/s. Then, a current of roughly J$=0.5$ A can be created around a ring of length roughly equal to r$=$100 nm. The magnetic field resulting by the ring current is equal to  $\mu_{0}$J$/(2\pi r)$, which is roughly equal to 1 Tesla. Hence, our concept can be implemented to produce large magnetic fields confined below 100 nm.

\section{Outlook}
An important application of the  concept we theoretically develop in this work is to implement it to optically create strong electron ring currents resulting in large magnetic fields of the order of Tesla. 
  To do so, one can envision using a spatio-temporal light spring, which is a pulse recently proposed \cite{ref:Pariente}. 
  Generating  light springs and understanding their nonlinear interaction with matter, such as plasmas to produce   beams of relativistic particles, is  of intense interest \cite{ref:Denoeud,ref:Shi,ref:Katoh,ref:Vieira1,ref:Vieira2}. 
  
  Light springs are exotic pulses resulting from a superposition of several different frequency Laguerre-Gauss orbital angular momentum (OAM) beams. Both their wavefront as well as their intensity profile  have a helical or corkscrew-like structure. Using high harmonic generation, it will be possible in the future to generate a focused down to roughly 100 nm VUV light spring from an OAM IR laser field.  
  Indeed, it was recently reported that roughly 12 eV VUV beams were generated by high harmonics from the MgO solid focused down to 100 nm \cite{ref:Korobenko}. The circular IR laser beam will be obtained from the fundamental IR laser field as well. The idea is that the corkscrew  intensity profile of the VUV light spring will release electrons at different points on a ring that is perpendicular to its propagation direction and at different times in the circular IR pulse.  This is equivalent to releasing electrons at different points on the ring with different phase delays between the VUV and IR pulses. Hence, a ring current can be generated to produce a large magnetic field along the propagation direction.

\section{Conclusions}
In conclusion, we have demonstrated a one-to-one mapping between the direction of ionization of an electron and the phase delay between a linearly polarized VUV pulse and a circular IR laser field. An ultra-short VUV pulse focused down to 100 nm or less releases the electron in the circular IR pulse with temporal and spatial resolution. Following release, the electron is then accelerated to high velocities by the IR pulse. 
 
 We have demonstrated this concept in the context of the N$_{2}$ molecule. However, future experiments can employ equally well atoms such as Helium, Argon or Neon. Selection of an atom for experiments should be partly based on the maximum IR intensity that can be considered without tunnel ionization of a valence electron. High IR intensities result in high electron speeds and thus large electron currents. For instance, the first ionization energy of Helium is higher than the first ionization energy of N$_{2}$. Hence, for Helium, intensities of the IR pulse higher  than  5$\times 10^{13}$ W/cm$^2$ can be considered, while  still keeping the rate of ionization of the valence electron via tunnelling  very small.  For experiments, selection of an atom should also be based on transitions from a valence or inner valence shell with a VUV pulse of around 20 eV photon energy  having large cross sections.
 
 The theoretical concept of control of electron motion developed here can be implemented to create  high magnetic fields. 
In the near future, it should be possible to generate focused VUV pulses with a corkscrew-like intensity profile that in conjunction with a circular IR pulse can direct electrons around a ring and create magnetic fields. However, the concept of controlling the direction of electron ionization with a VUV and IR pulse we theoretically develop here  is general and not restricted to creating ring currents. For instance, it can also be applicable to processes in physical chemistry, an area where coherent control  emerged as a tool to steer 
 a system into a particular final state \cite{ShaBru}. For instance, controlling the direction of ionization of an electron released from an inner valence or core orbital can influence from which atomic center  a valence electron is removed to fill in the hole created by the VUV pulse in a process known as interatomic Coulombic decay \cite{ref:Cederbaum}.

  \section{Acknowledgements}
A. E. wishes to thank Paul Corkum for spending one year of her sabbatical during 2020--2021 at the Joint Attosecond Science Lab of the National Research Council and the University of Ottawa. During this time, the idea for this work started. 
A. E. also thanks Matthias Kling, since part of the theoretical framework used in the current work was developed during a sabbatical in spring of 2021 in his group, at the time at Ludwig-Maximillian University in 
Munich. The authors acknowledge the use of the UCL Myriad High Throughput Computing Facility (Myriad@UCL), and associated support services, in the completion of this work. A.E. acknowledge the Leverhulme Trust Research Project Grant No. 2017-376. A.E. and G.P.K. acknowledge the EPSRC Grant EP/W005352/1. M. M. acknowledges funding from the EPSRC project 2419551. P. B. C. acknowledges support from the US Army Research Office under award number FA9550-16-1-0109 with contributions from the Canada Foundation for Innovation, the Canada Research Chairs program, Canada's Natural Sciences and Engineering Research Council and the National Research Council of Canada.


\newpage
\bibliography{bibliography}{}

\end{document}